\documentclass{article} %
\usepackage{iclr2027_conference,times}

\usepackage{amsmath,amsfonts,bm}

\def\eqref#1{equation~\ref{#1}}

\def\1{\bm{1}}

\DeclareMathAlphabet{\mathsfit}{\encodingdefault}{\sfdefault}{m}{sl}
\SetMathAlphabet{\mathsfit}{bold}{\encodingdefault}{\sfdefault}{bx}{n}

\usepackage{hyperref}
\usepackage{url}
\usepackage{xspace}
\usepackage{booktabs}
\usepackage{array}
\usepackage{multirow}
\usepackage{graphicx}
\usepackage{amsmath}
\usepackage{pifont}
\usepackage[table]{xcolor}
\usepackage{tikz}
\usepackage{pgfplots}
\pgfplotsset{compat=1.17}
\usetikzlibrary{arrows.meta, positioning, fit, backgrounds, calc}
\usepackage{listings}
\newcommand{\parabf}[1]{\vspace{0.35em}\noindent\textbf{#1}}
\definecolor{color-a}{RGB}{244, 241, 222}   %
\definecolor{color-b}{RGB}{129, 178, 154}   %
\definecolor{color-c}{RGB}{61, 64, 91}      %
\definecolor{color-d}{RGB}{242, 204, 143}   %
\definecolor{color-e}{RGB}{224, 122, 95}    %
\definecolor{color-f}{RGB}{201, 228, 202}   %
\definecolor{color-g}{RGB}{254, 217, 183}   %
\newcommand{\bench}{\textsc{VulContextBench}\xspace}

\title{\bench: A Benchmark for Security Context Retrieval in Coding Agents}

\author{%
Yikun Li\textsuperscript{1,}\thanks{Equal contribution.} \quad
Jinfeng Jiang\textsuperscript{1,*} \quad
Yieh Yuheng\textsuperscript{1} \quad
Ting Zhang\textsuperscript{2}\\
\bfseries Yide Yin\textsuperscript{3} \quad
Leow Wen Bin\textsuperscript{3} \quad
Eng Lieh Ouh\textsuperscript{1} \quad
Lwin Khin Shar\textsuperscript{1} \quad
David Lo\textsuperscript{1}\\[4pt]
{\normalfont\small\textsuperscript{1} Singapore Management University \quad
\textsuperscript{2} Monash University \quad \textsuperscript{3} GovTech}}

\hypersetup{%
  pdftitle={VulContextBench: A Benchmark for Security Context Retrieval in Coding Agents},
  pdfauthor={Yikun Li, Jinfeng Jiang, Yieh Yuheng, Ting Zhang, Yide Yin, Leow Wen Bin, Eng Lieh Ouh, Lwin Khin Shar, David Lo}}

\iclrfinalcopy %
\begin{document}

\maketitle
\fancyhead{}
\renewcommand{\headrulewidth}{0pt}

\begin{abstract}
Vulnerability-detection benchmarks score the verdict an agent reaches, not the evidence it gathered.
A model that recalls a CVE from pretraining therefore scores the same as one that traced the data flow.
We study a task where this difference matters, deciding whether a commit introduces a vulnerability.
Instead of scoring the verdict, we score whether the agent retrieved the code its conclusion depends on.
We present \bench, a benchmark of 111 vulnerability-introducing commits (VICs) across 83 repositories, 63 CWEs, and five languages.
Existing datasets label such commits by tracing a fix back through the version history, which often points to the wrong commit.
We therefore audit every case by hand against an explicit four-criterion definition of a VIC, so the benchmark does not inherit that label noise.
Each case is annotated with \emph{gold context}, 464 code blocks in total, each tagged by its role in the evidence for the vulnerability.
We evaluate seven frontier models with precision, recall and F1 at three granularities (file, block, and line), scored separately on the context an agent \emph{viewed} while exploring and on the context it finally \emph{declared} as evidence.
The gap between the two is the main finding.
Every model opens most of the gold context while exploring, but reports only part of it as evidence.
At the level of code blocks, the share of the gold context a model reports is 37 to 73 percentage points below the share it viewed.
Qwen3-Coder-Next views 86.3\% of the lines in annotated code blocks but cites only 12.9\% in its final report.
GPT-5.5, which cites the most, views 73\% and reports 36\%.
These results highlight a gap between finding relevant code and selecting it for the final report, which verdict-level benchmarks cannot reveal.

\end{abstract}

\section{Introduction}
\label{sec:intro}

LLM coding agents now operate at repository scale, exploring unfamiliar codebases with tools over many steps \citep{jimenez2024swebench, yang2024sweagent, dong2025survey}.
A natural application is security triage, where the agent is given a commit and must decide whether it introduces a vulnerability.
In this setting the usual way of evaluating agents has a specific problem.
Vulnerabilities carry public identifiers, public advisories, and public fixes, all of which appear in pretraining data, so a model can give the right verdict without ever having read the code.

\parabf{Limitations of Existing Benchmarks.}
Vulnerability-detection benchmarks present a function or a commit and score a binary label \citep{zhou2019devign, fan2020bigvul, perl2015vccfinder}.
Like the end-to-end \texttt{Pass@k} of issue-resolution benchmarks, this measures whether a system reaches the answer and ignores how it got there.
Two problems follow.
First, an outcome metric cannot separate reasoning from memorisation, and reported performance on these benchmarks drops sharply once labelling and duplication are handled realistically \citep{chakraborty2022deep, croft2023data, ding2025primevul}.
Models often succeed for reasons unrelated to the vulnerability \citep{steenhoek2023empirical}.
Second, and specific to the vulnerability-introduction setting, the labels come from automated tracing tools such as V-SZZ \citep{sliwerski2005changes, bao2022vszz}.
Some datasets check these labels by hand and others release them unchecked, so a labelled commit may not be the one that introduced the vulnerability.
Existing benchmarks therefore neither measure whether an agent retrieved the code its conclusion depends on, nor guarantee that the ground truth it is scored against is correct.

\begin{figure}[t]
\centering
\includegraphics[width=\linewidth]{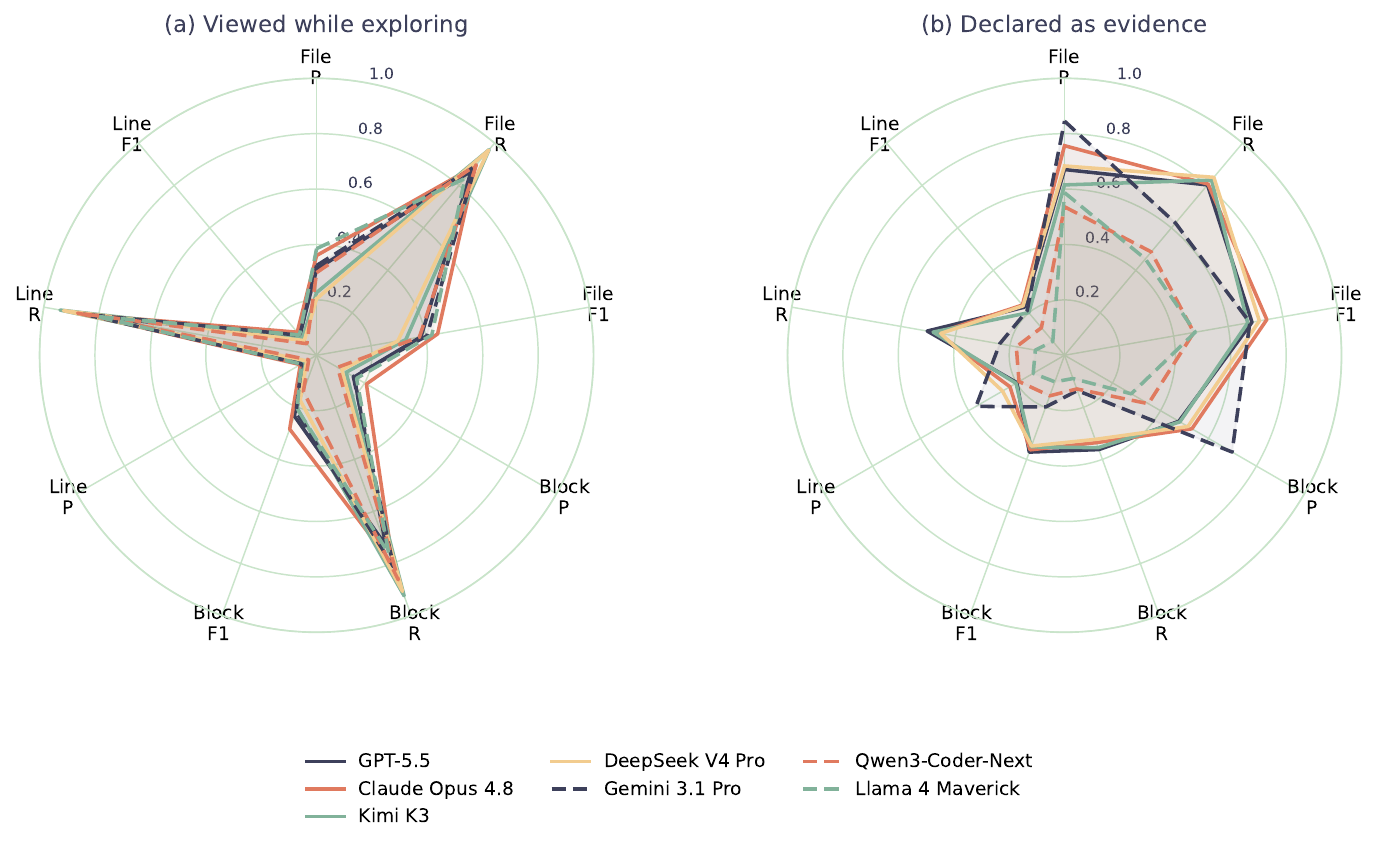}
\caption{\textbf{Agents find security-critical context, then fail to report it.}
Precision, recall and F1 at three granularities (file, block, and line) for seven models on the 111-task gold set.
\textbf{(a)} Context the agent \emph{viewed} while exploring: recall is high on every axis and precision is low, because agents open far more code than the gold contains, and the models almost overlap.
\textbf{(b)} Context the agent finally \emph{declared} as evidence: recall decreases while precision generally increases, and differences between models become clearer.
Much of the context retrieved in (a) is absent from the final citations, and the amount cited differs across models.
Panel (b) also shows the precision--recall trade-off.
Gemini 3.1 Pro has the highest precision on every axis and among the lowest recall, while Kimi K3 explores the most and has the second-highest declared recall at mid-range precision.}
\label{fig:radar}
\vspace{-5mm}
\end{figure}

\begin{figure}[t]
\centering
\includegraphics[width=\linewidth,trim={0bp 84bp 0bp 83bp},clip]{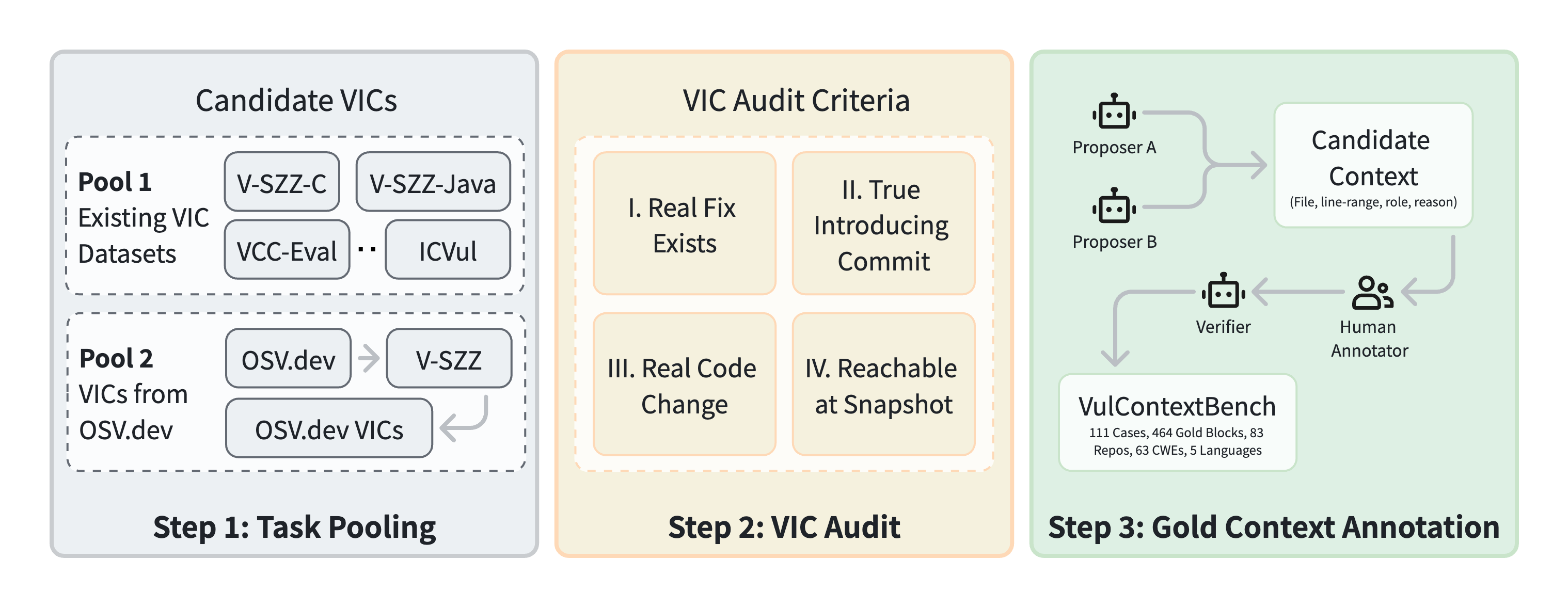}
\caption{\textbf{The \bench construction pipeline.}
Candidate VICs are pooled from six existing datasets and from OSV.dev advisories traced with V-SZZ, audited against four criteria using the repository history, and, if they pass, annotated with role-tagged gold context.
Most candidates fail the audit, so the corpus is 111 cases out of 307 candidates.}
\label{fig:pipeline}
\vspace{-3mm}
\end{figure}

\parabf{\bench: Beyond Verdict-Level Evaluation.}
We build \bench through a three-step pipeline (\autoref{fig:pipeline}).
\textbf{(1)~Task pooling}: we use candidate vulnerability-introducing commits (VICs) from six existing datasets, both human-curated and V-SZZ-labelled.
We add a second pool of recent VICs, traced from OSV.dev advisories with V-SZZ \citep{bao2022vszz}, because the existing VIC datasets end in 2024.
\textbf{(2)~VIC audit}: every candidate is judged against four explicit criteria: a real CVE with a fixing commit; the commit is the true introducer rather than the last to touch the vulnerable line; the commit really changed the vulnerable code rather than, for example, only reformatting it or importing the whole project; and the vulnerability is reachable by an attacker at that commit.
Each is checked against the full repository history.
\textbf{(3)~Gold context annotation}: for each surviving case, two proposer agents draft candidate context with full knowledge of the case.
An annotator reviews every block and writes the gold set as line ranges in the snapshot.
The result is 111 cases carrying 464 gold blocks over 13{,}370 lines, across five languages, each block tagged with its role in the evidence for the vulnerability (VIC core, sink, reachability, dependency).

We score an agent by comparing the code it retrieved with the gold context.
The comparison is made at three granularities: \emph{file}, \emph{block} (the function or region that contains the vulnerable code), and \emph{line} (the exact lines the fix changes).
Each agent is scored twice: on everything it \emph{viewed} while exploring, and on the blocks it finally \emph{declared} as evidence.
We evaluate seven frontier models: GPT-5.5, Claude Opus 4.8, Kimi K3, DeepSeek V4 Pro, Gemini 3.1 Pro, Qwen3-Coder-Next and Llama 4 Maverick.
Each receives only the repository snapshot and the introducing diff, with no CVE or CWE identifier, and must explore the repository and then declare the blocks it considers necessary.

\parabf{Takeaways.}
\ding{182}~\textbf{Agents omit relevant code from their final citations.}
All seven models inspect much more of the annotated evidence during exploration than they cite in their final reports.
For example, Qwen3-Coder-Next views 86.3\% of the lines in annotated code blocks but cites only 12.9\% (\autoref{tab:retrieval}).
This gap shows why we evaluate both repository exploration and the evidence an agent presents to the user.
\ding{183}~\textbf{Citing the right files does not mean citing the right code within them.}
When we check only which files agents cite, GPT-5.5 and Gemini 3.1 Pro score similarly (F1 0.687 and 0.679).
When we check the cited code within relevant functions or regions, called blocks, their scores differ substantially (F1 0.372 and 0.199; \autoref{tab:retrieval}(b)).
Evaluating citations within files therefore reveals differences that file-level scores hide.
\ding{184}~\textbf{Agents struggle to cite and identify supporting evidence.}
We distinguish four evidence roles: vulnerability-introducing code, harmful operations, attacker reachability, and supporting declarations.
All seven models have higher recall for vulnerability-introducing code than for the three supporting roles.
With correct role labels required, mean recall for vulnerability-introducing code is more than three times that for supporting declarations (\autoref{tab:roles}).

\parabf{Contributions.}
\begin{itemize}
\item \textbf{A task formulation and evaluation protocol} for security context retrieval (Section~\ref{sec:bench}): gold context with role-tagged blocks, precision/recall/F1 at file, block and line granularity, scored over viewed and declared context separately, against gold context that a blinded verifier confirmed is sufficient.
\item \textbf{The \bench corpus and baselines} (Sections~\ref{sec:bench} and~\ref{sec:experiments}): 111 audited cases across five languages with 464 role-tagged gold blocks, and an evaluation of seven frontier models under all three granularities and both context stages.
\end{itemize}

\section{\bench: A Benchmark for Security Context Retrieval}
\label{sec:bench}

\subsection{Task Formulation}
\label{sec:task}

An instance is a pair $(\mathcal{S}, d_{\mathrm{vic}})$: a read-only snapshot $\mathcal{S}$ of the repository at the candidate commit and the introducing diff $d_{\mathrm{vic}}$.
The agent receives nothing else: no CVE identifier, no description, no CWE identifier, and no fixing commit.
The agent explores $\mathcal{S}$ and returns blocks
\[
C^{A} = \{(f_i, s_i, e_i, c_i, r_i)\}_{i=1}^{n},
\]
where $f_i$ is a path in $\mathcal{S}$, $[s_i,e_i]$ an inclusive line range, $c_i$ a category, and $r_i$ a one-sentence justification.
For role-specific evaluation, submitted labels are mapped to four benchmark categories: the \emph{VIC core} (the lines that introduce the vulnerability), the \emph{sink} (where the flaw becomes harmful), \emph{reachability} (the path from an attacker-controlled entry point to the vulnerable code), and the \emph{dependency} (the definition, invariant, or configuration elsewhere that the changed code depends on).
Only the last three are context in the strict sense.
The VIC core is the vulnerability itself, and we score it as a fourth role because locating it is a retrieval problem in its own right.
\emph{Gold context} refers to the union of all four.

\subsection{Step 1: Task Pooling}
\label{sec:step1}

Candidates come from two pools, each included for a different reason.

\parabf{Pool 1: existing VIC datasets.}
We pooled candidates from six published sources of two kinds.
\emph{Human-curated}: V-SZZ-C and V-SZZ-Java, where the source authors checked the introducing commit for each CVE by hand \citep{bao2022vszz}; VCC-Eval, a tool-assisted human review of Java vulnerability-contributing commits following \citet{perl2015vccfinder}; and the Vulnerability History Project \citep{vhp} at its highest review level.
\emph{V-SZZ-labelled}: ICVul for C/C++ \citep{lu2025icvul} and the Java dataset of \citet{aladics2022vic}, both traced with V-SZZ over curated fix-commit collections \citep{ponta2019projectkb}.
We take the labels of these datasets as published, without re-running V-SZZ ourselves, and remove duplicate CVEs.

\parabf{Pool 2: recent VICs traced from OSV.dev.}
The published datasets have two limits.
They cover only C and Java; we found no comparable CVE-tagged VIC dataset for other languages.
They are also not current, since the newest of the six ends in 2024 and the human-curated ones in 2022.
A benchmark built only from them would consist entirely of vulnerabilities whose advisories, fixes and discussions have been public for years and may have been seen during model training.
For such a benchmark, memorisation is an alternative explanation for any result.
We therefore built a second pool ourselves.
From OSV.dev we took advisories for CVEs assigned in 2026 that reference a fixing commit, across the PyPI, npm, Go and Maven ecosystems.
For each advisory, we applied V-SZZ \citep{bao2022vszz} to trace the VIC of each vulnerability.

\subsection{Step 2: VIC Audit}
\label{sec:step2}

\parabf{Why every candidate is audited.}
The labels in both pools were produced by V-SZZ or checked by hand, in Pool 2 by our own run of V-SZZ.
V-SZZ traces the lines a fix changes back to the commit that introduced them, which is not always the commit that made the code vulnerable.
Some upstream datasets check their labels by hand and others release them unchecked.
Prior evaluations of the tools V-SZZ builds on against ground truth confirmed by developers found large disagreement \citep{rosa2021evaluating}.
A benchmark that trusted these labels would score agents against ground truth that is mostly wrong.
We therefore audit every candidate against four criteria, each checked with evidence from the repository history.
The audit keeps 111 of the 307 candidates that enter it.
Measuring the accuracy of upstream labelling is not the subject of this paper.
The audit exists so that the benchmark does not inherit its errors.

\parabf{The four criteria.}
A commit is a VIC only if it satisfies all four.
\textbf{(i) A real vulnerability with a real fix}: a CVE exists and a vulnerability-fixing commit (VFC) removes it.
The fix defines where the vulnerability is, so without a trustworthy fix the label cannot be checked.
\textbf{(ii) The true introducer, not the last commit to touch it}: in the full repository history, the lines the VFC changes must trace back to this commit, and the vulnerable behaviour must not exist before it.
A commit that only reformatted or moved already vulnerable code does not qualify, whatever its own diff shows.
This is the most common failure, because the label points to whichever commit last edited the line.
\textbf{(iii) A real change to the vulnerable code}: the commit is not a rename, a reformat, a whole-project import or move, an earlier incomplete fix, or a commit that touches other files than the vulnerable code.
Such a commit is the last to touch every line it imported or reformatted, so it is labelled as the introducer while introducing nothing.
\textbf{(iv) Reachable at that snapshot}: the vulnerability is attacker-reachable at the candidate commit, not a latent flaw whose entry point or sink is added in a later commit.
The agent is given the snapshot at this commit, so a vulnerability that is not yet reachable there has no reachability context to retrieve, and the task would be unanswerable.

\subsection{Step 3: Gold Context Annotation}
\label{sec:step3}

For each surviving case, two independent proposers draft candidate context: a Claude Code agent running Claude Opus 4.8 and an OpenAI Codex agent running GPT-5.5.
Unlike the agents we later evaluate, the proposers are not blinded.
Each receives everything we know about the case: the repository snapshot, the introducing diff, the fixing diff, and the CVE identifier, CWE and description.
The purpose of this step is to find the right context, not to test whether an agent can find it.
The two proposers work in isolated sandboxes and cannot see each other's output.
A deterministic diff, based on the line-range overlap of the two proposals, then separates the blocks both proposers found from the blocks only one of them found.
The annotator reviews every block in both sets, checks each against the repository, and writes the gold set as exact line ranges in the snapshot.
Blocks are emitted at the granularity of the enclosing syntactic unit: a function, method, class, struct, macro, top-level statement, or an explicitly labelled file section when the relevant code is not inside a function.
Each line in a block is marked as authored by the commit or not.

\parabf{A blinded test that the gold context is enough.}
We independently assess the sufficiency of the gold context.
We test, under blinding and against a control, whether it is enough to establish the vulnerability.
A separate model, GPT-5.5, serves as the verifier $V$ and is given two things and nothing else: the introducing diff $d_{\mathrm{vic}}$ and the text of the gold context $C^{G}$.
It is asked whether the commit introduces a vulnerability and what that vulnerability is.
We then run the same verifier with only the diff, without the gold blocks.
The gold context is \textbf{sufficient}, over the corpus as a whole, if the verifier identifies the vulnerability more often with the gold blocks than without them.
Over the 111 cases, GPT-5.5 identifies the vulnerability in 27.0\% of samples with the diff alone and in 75.1\% with the gold context, a lift of 48 points.
The verifier never sees the CVE description, the CVE identifier, the CWE identifier, the fixing commit, or the repository.
Withholding the CVE description is the key design choice.
The description names the vulnerability and the CWE identifier names its class, so a verifier that saw either could identify the vulnerability without reading any code, and we could no longer tell whether the gold blocks helped.

\subsection{\bench Statistics}
\label{sec:stats}

\autoref{tab:overview} summarises the corpus.
The 111 cases come from 83 repositories, and the largest contributes 7 cases, which limits the repository-level leakage that inflates results on datasets with many commits per project \citep{croft2023data}.
CVE years span 2008 to 2026, with 35 cases from 2024 and 34 from 2026, over 63 distinct CWEs, the most frequent being CWE-22 (11), CWE-125 (6), and CWE-416 (5).
The 2026 cases are recent enough to limit how much of the corpus can have been absorbed into model pretraining.

\begin{table}[t]
\centering
\footnotesize
\caption{\textbf{Statistics of \bench.}
111 audited VICs annotated with role-tagged gold context.
\textbf{\# Files} counts annotated files summed over tasks, \textbf{\# Blocks} the gold blocks, and \textbf{\# Lines} the lines those blocks span.
Per-language CWE counts overlap, so the total is the union.}
\label{tab:overview}
\begin{tabular}{lrrrrrr}
\toprule
\multirow{2}{*}{\textbf{Language}} & \multirow{2}{*}{\textbf{\# Repos}} &
\multirow{2}{*}{\textbf{\# Tasks}} & \multirow{2}{*}{\textbf{\# CWEs}} &
\multicolumn{3}{c}{\textbf{Context Statistics}} \\
\cmidrule(lr){5-7}
 & & & & \textbf{\# Files} & \textbf{\# Blocks} & \textbf{\# Lines} \\
\midrule
C / C++     & 31 & 54 & 33 & 101 & 197 & 7{,}698 \\
Java        & 25 & 28 & 19 & 74 & 125 & 2{,}654 \\
JavaScript  & 9 & 10 & 9 & 17 & 49 & 1{,}182 \\
Python      & 9 & 10 & 8 & 21 & 48 & 923 \\
Go          & 9 & 9 & 8 & 29 & 45 & 913 \\
\midrule
\rowcolor{color-a}
\textbf{Total} & \textbf{83} & \textbf{111} & \textbf{63} & \textbf{242} & \textbf{464} & \textbf{13{,}370} \\
\bottomrule
\end{tabular}
\end{table}

\subsection{Evaluation Protocol}
\label{sec:protocol}

Let $C^{G}$ be the gold context and $C^{A}$ the agent's context, each a set of $(\text{file}, \text{line-range})$ regions.
We align them and report precision, recall and F1 at three granularities, chosen so that each isolates a different failure:

\begin{itemize}
\item \textbf{File}: sets of file paths.
Did the agent look in the right places?
Coarse enough that most models pass.
\item \textbf{Block}: the enclosing syntactic unit, meaning the function, method, or region containing the vulnerability.
Did the agent isolate the code that carries the evidence, without including the rest of the file?
\item \textbf{Line}: the exact lines the fixing commit changes.
Did the agent locate the flaw itself?
Precision is low here by construction, since agents return whole blocks and the vulnerability is a few lines inside one.
Recall is the informative measure.
\end{itemize}

Writing $I = |C^{A} \cap C^{G}|$ for the overlap, precision is $P = I/|C^{A}|$, recall is $R = I/|C^{G}|$, and $F_1 = 2PR/(P+R)$ is their harmonic mean.
Scores are macro-averaged over tasks.
Precision matters here for a reason specific to security review.
Every retrieved block that is not part of the evidence is a block a reviewer has to read and rule out.
Returning too much also makes it harder for a downstream model to use what was retrieved \citep{liu2024lost}.

\parabf{Viewed versus declared.}
We capture two context sets per run.
The \emph{viewed} set is every region the agent opened during exploration, recovered from its trajectory.
The \emph{declared} set is the block list it finally submits as the evidence for its conclusion.
Scoring both separates two abilities that a single number combines, finding the code and recognising which of it mattered.

\section{Experiments}
\label{sec:experiments}

\parabf{Setup.}
We evaluate seven models over the full 111-task gold set: GPT-5.5, Claude Opus 4.8, Kimi K3, DeepSeek V4 Pro, Gemini 3.1 Pro, Qwen3-Coder-Next and Llama 4 Maverick.
Each is given the repository snapshot at the candidate commit and the introducing diff, and must explore and then declare the blocks it considers necessary.
All seven run in the same harness, mini-SWE-agent \citep{lieret2025miniswe}, with the same prompt.

Each agent produces a structured security report through a final submission tool.
The report includes a verdict, a written explanation of the root cause and security impact, and an evidence list.
Each evidence entry specifies a repository-relative file path, start and end lines, a role, and a brief justification.
The prompt instructs agents to cite only inspected code and verify citation boundaries using numbered source output.
We take these cited ranges as declared evidence and recover viewed context separately from the exploration trace.

\subsection{Benchmarking LLMs on Security Context Retrieval}

\autoref{tab:retrieval} reports the seven models on both context sets at the three granularities of Section~\ref{sec:protocol}.
GPT-5.5 leads on declared block F1 (0.372), with Claude Opus 4.8 (0.363), Kimi K3 (0.352) and DeepSeek V4 Pro (0.349) close behind and Llama 4 Maverick last (0.102).
The separation between the models depends on the panel and on the granularity.

\parabf{Exploration separates the models far less than declaration.}
In panel~(a), block recall ranges from 0.731 to 0.924 and line recall from 0.768 to 0.941, so the widest explorer views only about 1.3 times as much of the gold as the narrowest.
In panel~(b), the best model declares 4.1 times as much of the gold as the worst (block recall 0.362 against 0.089).
Even the weakest explorer opens most of the code the gold context is drawn from.
Precision is low for every model (0.087 to 0.208 at block level) because agents open far more code than the gold contains.
This is expected rather than a failure, because an agent cannot know in advance which of the files it opens will matter.

\begin{table}[t]
\centering
\footnotesize
\caption{\textbf{Retrieval at three granularities}, 111-task gold set, macro-averaged.
\textbf{(a)} every region the agent opened while exploring; \textbf{(b)} the blocks it finally submitted as evidence.
Exploration recall is high and close across models, and precision is low because agents open far more code than the gold contains.
Model scores are similar in (a) and differ more at block and line granularity in (b).
Verdict accuracy, agreement with the audited label, is a property of the model rather than of a context set, so it is repeated in both panels for reference.
The best value in each column is shaded and bold.}
\label{tab:retrieval}
\setlength{\tabcolsep}{4.2pt}

\textbf{(a) Explored Context}\\[0.25em]
\begin{tabular}{lcccccccccc}
\toprule
\multirow{2}{*}{\textbf{Model}} &
\multicolumn{3}{c}{\textbf{File}} & \multicolumn{3}{c}{\textbf{Block}} &
\multicolumn{3}{c}{\textbf{Line}} & \multirow{2}{*}{\shortstack{\textbf{Verdict}\\\textbf{Accuracy}}} \\
\cmidrule(lr){2-4}\cmidrule(lr){5-7}\cmidrule(lr){8-10}
 & P & R & F1 & P & R & F1 & P & R & F1 & \\
\midrule
GPT-5.5               & 0.310 & 0.868 & 0.383 & 0.157 & 0.731 & 0.234 & 0.057 & 0.768 & 0.094 & \cellcolor{color-f}\textbf{0.766} \\
Claude Opus 4.8       & 0.360 & 0.892 & \cellcolor{color-f}\textbf{0.444} & \cellcolor{color-f}\textbf{0.208} & 0.808 & \cellcolor{color-f}\textbf{0.284} & \cellcolor{color-f}\textbf{0.068} & 0.843 & \cellcolor{color-f}\textbf{0.109} & 0.622 \\
Kimi K3               & 0.226 & \cellcolor{color-f}\textbf{0.969} & 0.327 & 0.123 & \cellcolor{color-f}\textbf{0.924} & 0.204 & 0.047 & \cellcolor{color-f}\textbf{0.941} & 0.079 & 0.555 \\
DeepSeek V4 Pro       & 0.204 & 0.966 & 0.298 & 0.105 & 0.907 & 0.173 & 0.041 & 0.928 & 0.069 & 0.725 \\
Gemini 3.1 Pro        & 0.324 & 0.891 & 0.404 & 0.153 & 0.814 & 0.216 & 0.062 & 0.854 & 0.094 & 0.748 \\
Qwen3-Coder-Next      & 0.298 & 0.897 & 0.377 & 0.087 & 0.863 & 0.139 & 0.033 & 0.876 & 0.055 & 0.469 \\
Llama 4 Maverick      & \cellcolor{color-f}\textbf{0.385} & 0.833 & 0.423 & 0.167 & 0.761 & 0.206 & 0.064 & 0.810 & 0.096 & 0.171 \\
\bottomrule
\end{tabular}

\vspace{0.6em}
\textbf{(b) Declared Evidence}\\[0.25em]
\begin{tabular}{lcccccccccc}
\toprule
\multirow{2}{*}{\textbf{Model}} &
\multicolumn{3}{c}{\textbf{File}} & \multicolumn{3}{c}{\textbf{Block}} &
\multicolumn{3}{c}{\textbf{Line}} & \multirow{2}{*}{\shortstack{\textbf{Verdict}\\\textbf{Accuracy}}} \\
\cmidrule(lr){2-4}\cmidrule(lr){5-7}\cmidrule(lr){8-10}
 & P & R & F1 & P & R & F1 & P & R & F1 & \\
\midrule
GPT-5.5               & 0.671 & 0.803 & 0.687 & 0.477 & \cellcolor{color-f}\textbf{0.362} & \cellcolor{color-f}\textbf{0.372} & 0.199 & \cellcolor{color-f}\textbf{0.503} & 0.226 & \cellcolor{color-f}\textbf{0.766} \\
Claude Opus 4.8       & 0.757 & 0.807 & \cellcolor{color-f}\textbf{0.742} & 0.531 & 0.336 & 0.363 & 0.228 & 0.465 & \cellcolor{color-f}\textbf{0.237} & 0.622 \\
Kimi K3               & 0.615 & 0.826 & 0.671 & 0.481 & 0.355 & 0.352 & 0.202 & 0.482 & 0.201 & 0.555 \\
DeepSeek V4 Pro       & 0.683 & \cellcolor{color-f}\textbf{0.839} & 0.714 & 0.514 & 0.323 & 0.349 & 0.258 & 0.454 & 0.233 & 0.725 \\
Gemini 3.1 Pro        & \cellcolor{color-f}\textbf{0.845} & 0.623 & 0.679 & \cellcolor{color-f}\textbf{0.698} & 0.136 & 0.199 & \cellcolor{color-f}\textbf{0.369} & 0.238 & 0.210 & 0.748 \\
Qwen3-Coder-Next      & 0.536 & 0.487 & 0.474 & 0.350 & 0.129 & 0.156 & 0.190 & 0.177 & 0.130 & 0.469 \\
Llama 4 Maverick      & 0.587 & 0.454 & 0.479 & 0.278 & 0.089 & 0.102 & 0.131 & 0.107 & 0.066 & 0.171 \\
\bottomrule
\end{tabular}
\end{table}

\parabf{Every model cites only part of what it found.}
Comparing the two panels gives the paper's main result.
On the same tasks, \emph{every} model declares far less of the gold context than it had already opened.
DeepSeek V4 Pro views 0.907 of the gold block lines and declares 0.323 of them; Qwen3-Coder-Next views 0.863 and declares 0.129; even GPT-5.5, which cites the most, falls from 0.731 to 0.362.
The gap ranges from 37 to 73 points of block recall, and is largest for the lowest-scoring systems.
The declared ranking has little relation to the explored one.
Kimi K3, DeepSeek V4 Pro and Qwen3-Coder-Next explore the most (0.924, 0.907 and 0.863 block recall) and they place third, fourth and sixth of seven on declared block F1.
The models find most of the code, and the difference between them is in what they select for citation.
The same result has been reported for issue resolution \citep{li2026contextbench}.
We find that it also holds for security, and that it is larger here.

\parabf{Exploring more does not mean declaring more.}
The share of viewed gold block lines included in the declared evidence ranges from 0.12 to 0.50 across the seven models, and it is unrelated to how much a model explores.
GPT-5.5 explores the least at block and line level and declares the most, citing about half of the gold block lines it viewed.
The three models that explore the most all cite less.
What separates the models is how much of what they opened they cite as evidence, not how much they opened.

\parabf{Finer-grained scores reveal differences that file-level scores can hide.}
Declared evidence F1 ranges from 0.47 to 0.74 at file level, 0.10 to 0.37 at block level, and 0.07 to 0.24 at line level.
The absolute range does not consistently increase at finer granularities, but models with similar file-level scores can differ substantially in their coverage of code within those files.
For example, GPT-5.5 and Gemini 3.1 Pro have similar file F1 (0.687 and 0.679), while GPT-5.5 has substantially higher block F1 (0.372 against 0.199).
Scoring below the file level therefore distinguishes citing relevant files from identifying the relevant code within them.

\parabf{Precision and recall rank the models differently, and either can serve the verdict.}
The two models with the highest verdict accuracy differ in opposite directions.
GPT-5.5 leads on declared recall (0.362), and Gemini 3.1 Pro has the highest declared precision at every granularity (0.698 at block level) with the third-lowest recall.
Kimi K3 declares more of the gold than Gemini 3.1 Pro (0.355 against 0.136) but reaches the correct verdict less often (0.555 against 0.748).
Neither precision nor recall alone determines the verdict.

\subsection{Which Kinds of Context Are Retrieved, and Which Are Not?}

Not all gold context is the same kind of code.
Each block is tagged with its role in the evidence for the vulnerability: the lines that introduce the vulnerability (the \emph{VIC core}, which is the flaw rather than context for it, as Section~\ref{sec:task} notes), the \emph{sink} where the flaw becomes harmful, the \emph{reachability} path from attacker-controlled input, and the \emph{dependency}, meaning not a package but the declaration elsewhere that the changed code depends on.
Because both the gold blocks and the agents' declared blocks carry a role, we can score each role on its own.
A declared block counts toward a role only if the agent tagged it as that role, and a gold block of a role is recalled only by a block tagged with that role.
\autoref{tab:roles} reports block-level precision and recall per role.

\parabf{Roles differ in recall, not in precision.}
On recall, the VIC core is the best role for all seven models and the dependency the worst for five of them.
Mean recall runs 0.408 for the VIC core, 0.227 for reachability, 0.166 for the sink and 0.115 for the dependency, a spread of more than three times.
Precision varies much less across roles: 0.513, 0.337, 0.328 and 0.243.
So what separates the roles is not how often an agent is right when it names one, but how much of the gold of that kind it reaches.
Models do best on the one role the diff partly gives them, since the VIC core is by construction the code the commit touched.
They do worst on the role that can only be found by working out \emph{why} the change is vulnerable.

\parabf{Much of the supporting context is found but mislabelled.}
Scoring the same blocks with the agents' tags ignored raises mean recall from 0.408 to 0.620 for the VIC core, from 0.166 to 0.354 for the sink, from 0.227 to 0.420 for reachability and from 0.115 to 0.237 for the dependency.
About a third of the VIC core blocks that agents cover, and about half of the sink, reachability and dependency blocks, are declared under some other role.
The per-role gap therefore has two parts, context that is not retrieved at all and context that is retrieved but not recognised for what it is.
For the three supporting roles, the second part accounts for about half of what agents find.

\parabf{Precision and recall move together, except on the dependency.}
Across the seven models, precision and recall are strongly correlated for the VIC core, the sink and reachability (Pearson $r=0.96$, $0.91$ and $0.78$).
A model that finds more blocks of a role also labels them correctly more often, so there is no trade-off between the two within a role.
For the dependency the two are unrelated ($r=0.04$).
Finding the declaration the changed code depends on and recognising it as such are separate abilities.
This fits the role's definition, since it is the one that requires understanding why the change is vulnerable rather than where it is.

\parabf{No single model leads on every role.}
On recall, DeepSeek V4 Pro leads on the VIC core (0.555) and the sink (0.311), GPT-5.5 on reachability (0.408) and Kimi K3 on the dependency (0.228).
On precision, Claude Opus 4.8 leads on the VIC core (0.651) and reachability (0.446) and Kimi K3 on the sink (0.471).
Gemini 3.1 Pro's dependency precision is computed from few tagged blocks and should not be read as a lead.

\begin{table}[t]
\centering
\scriptsize
\caption{\textbf{Which kinds of context do agents retrieve and label correctly?}
Block-level precision and recall per role over the 111-task gold set, where a declared block counts toward a role only if the agent tagged it with that role.
The best value in each column is shaded and bold, among the per-model rows only.}
\label{tab:roles}
\setlength{\tabcolsep}{5pt}
\begin{tabular}{lcccccccc}
\toprule
\multirow{2}{*}{\textbf{Model}} & \multicolumn{4}{c}{\textbf{Precision}} & \multicolumn{4}{c}{\textbf{Recall}} \\
\cmidrule(lr){2-5}\cmidrule(lr){6-9}
 & \textbf{VIC Core} & \textbf{Sink} & \textbf{Reachability} & \textbf{Dependency} & \textbf{VIC Core} & \textbf{Sink} & \textbf{Reachability} & \textbf{Dependency} \\
\midrule
GPT-5.5               & 0.570 & 0.438 & 0.346 & 0.250 & 0.461 & 0.273 & \cellcolor{color-f}\textbf{0.408} & 0.176 \\
Claude Opus 4.8       & \cellcolor{color-f}\textbf{0.651} & 0.348 & \cellcolor{color-f}\textbf{0.446} & 0.178 & 0.510 & 0.152 & 0.317 & 0.109 \\
Kimi K3               & 0.600 & \cellcolor{color-f}\textbf{0.471} & 0.428 & 0.271 & 0.450 & 0.266 & 0.279 & \cellcolor{color-f}\textbf{0.228} \\
DeepSeek V4 Pro       & 0.604 & 0.380 & 0.336 & 0.240 & \cellcolor{color-f}\textbf{0.555} & \cellcolor{color-f}\textbf{0.311} & 0.307 & 0.189 \\
Gemini 3.1 Pro        & 0.643 & 0.313 & 0.415 & \cellcolor{color-f}\textbf{0.444} & 0.460 & 0.096 & 0.221 & 0.038 \\
Qwen3-Coder-Next      & 0.378 & 0.182 & 0.219 & 0.153 & 0.285 & 0.043 & 0.041 & 0.048 \\
Llama 4 Maverick      & 0.145 & 0.167 & 0.167 & 0.167 & 0.137 & 0.021 & 0.014 & 0.016 \\
\midrule
\textbf{Mean} & \textbf{0.513} & \textbf{0.328} & \textbf{0.337} & \textbf{0.243} & \textbf{0.408} & \textbf{0.166} & \textbf{0.227} & \textbf{0.115} \\
\bottomrule
\end{tabular}
\end{table}

\section{Related Work}
\label{sec:related}

\parabf{Context retrieval for coding agents.}
Benchmarks for repository-scale coding agents mostly use outcome metrics.
SWE-bench \citep{jimenez2024swebench} and agent scaffolds built for it \citep{yang2024sweagent} score whether tests pass.
A recent line of work argues for also evaluating the intermediate context an agent retrieves, rather than only the outcome.
ContextBench \citep{li2026contextbench} annotates gold context for 1{,}136 issue-resolution tasks across 8 languages and scores agents by precision/recall/F1 at file, block, and line granularity.
SWE Context Bench \citep{zhu2026swecontextbench} evaluates whether agents reuse context across tasks, measured by accuracy, time, and cost against a no-context baseline.
We adopt the evaluation approach and metric structure of the former, and apply it to a setting where it is needed more.
An issue-resolution benchmark has an execution-based outcome signal (tests) that partly checks the end state, whereas no test fails because a commit introduced a CVE years ago.
In our setting, retrieval quality is therefore not an addition to the outcome signal.
It is the only process signal available.
Two further differences come from the domain.
Our gold blocks are \emph{role-tagged} (VIC core, sink, reachability, dependency), because the pieces of evidence for a vulnerability depend on one another.
Our ground truth also had to be audited before it could be annotated.
\autoref{tab:related} places \bench against both lines of work.

\begin{table}[htb]
\centering
\footnotesize
\caption{\textbf{Positioning.}
Vulnerability benchmarks score verdicts against labels they do not audit.
Context-retrieval benchmarks score process but target issue resolution, where labels are validated by running tests.}
\label{tab:related}
\begin{tabular}{llcccc}
\toprule
\textbf{Benchmark} & \textbf{Task} & \textbf{Scored Unit} & \textbf{Gold} & \textbf{Process} & \textbf{Labels} \\
 & & & \textbf{Context} & \textbf{Signal} & \textbf{Audited} \\
\midrule
Big-Vul & Vulnerability Detection & Function & \ding{55} & \ding{55} & \ding{55} \\
PrimeVul & Vulnerability Detection & Function & \ding{55} & \ding{55} & \ding{55} \\
SWE-bench & Issue Resolution & Patch & \ding{55} & \ding{55} & \ding{51} \\
ContextBench & Issue Resolution & Context & \ding{51} & \ding{51} & \ding{51} \\
SWE Context Bench & Context Reuse & Context & \ding{51} & \ding{51} & \ding{51} \\
\midrule
\rowcolor{color-a}
\textbf{\bench} & \textbf{Vulnerability Context} & \textbf{Context} & \ding{51} & \ding{51} & \ding{51} \\
\bottomrule
\end{tabular}
\end{table}

\parabf{V-SZZ and VIC identification.}
V-SZZ \citep{bao2022vszz} extends SZZ \citep{sliwerski2005changes} to vulnerabilities and traces a fix back through the version history to the commit that introduced the repaired lines.
Dataset pipelines of the V-SZZ family exist for Java \citep{aladics2022vic} and C/C++ \citep{lu2025icvul} over curated fix collections \citep{ponta2019projectkb}.
\citet{rosa2021evaluating} evaluated SZZ-family implementations against ground truth confirmed by developers and found large disagreement, for general bugs and in aggregate.
We ask the narrower question for vulnerabilities and answer it per case, auditing every candidate rather than taking it as labelled.

\parabf{Vulnerability datasets and label noise.}
Devign \citep{zhou2019devign}, Big-Vul \citep{fan2020bigvul}, and VCCFinder \citep{perl2015vccfinder} established commit- and function-level detection.
Subsequent studies identified label noise in vulnerability datasets \citep{li2024cleanvul,li2026out}.
\citet{chakraborty2022deep} showed accuracy dropping sharply on realistic data, \citet{jimenez2019importance} showed that results depend on labelling assumptions, \citet{croft2023data} measured accuracy, uniqueness and consistency problems, and PrimeVul \citep{ding2025primevul} showed reported performance falling once label noise and duplication are handled.
These findings motivate verification of the \emph{commit-level VIC label} for every case.
\citet{steenhoek2023empirical} also found that models often succeed for reasons unrelated to the vulnerability, which is the failure our metric is designed to detect.

\section{Conclusion}

We introduced \bench, which scores the code a coding agent retrieves when deciding whether a commit introduces a vulnerability, not only its verdict.
Its 111 cases are audited against four explicit criteria and carry role-tagged gold context that a blinded verifier confirms is sufficient.
Across seven frontier models, every model views most of the gold context but cites much less of it in the final report.
The models separate only at block and line granularity, and the supporting roles of context are recalled far less often than the vulnerable code itself.
None of this is visible to verdict-level evaluation.
A natural next step is proof-of-concept execution, so that reachability is confirmed by running an exploit rather than by reading the code.
The annotations also provide a basis for testing whether better evidence selection improves security reports and helps reviewers verify the claims agents make.

The benchmark is available in our replication package at \url{https://github.com/yikun-li/vul-context-bench}.

\newpage
\bibliography{references}
\bibliographystyle{iclr2027_conference}

\end{document}